\documentclass[11pt,twoside,lineno]{pnas-new}

\templatetype{pnasresearcharticle} 

\usepackage{comment}
\usepackage{dcolumn}
\newcolumntype{d}[1]{D{.}{.}{#1}}
\usepackage{bm}
\usepackage[capitalise,nameinlink]{cleveref}
\usepackage{chemformula}
\usepackage[disable]{todonotes}
\usepackage{multirow}
\usepackage{multicol}
\usepackage{natbib}
\usepackage[utf8]{inputenc}
\usepackage{setspace}
\usepackage{pdfpages}

\graphicspath{
	{./figures/}
}

\title{Neighborhoods and Functionality in Metals}

\author[a]{M.\ Rajivmoorthy}
\author[a]{T.\ R.\ Wilson}
\author[a,*]{M.\ E.\ Eberhart}

\affil[a]{Molecular Theory Group, Colorado School of Mines, Golden, Colorado, USA}
\affil[*]{To whom correspondence should be addressed. E-mail: \href{mailto:meberhar@mines.edu?subject=Neighborhoods\%20in\%Structural \%20Materials}{meberhar@mines.edu}}

\leadauthor{Rajivmoorthy}

\authordeclaration{The authors have no conflicts of interest to declare.}

\keywords{\textit{Keywords:} 
	 nearsightedness $|$ neighborhoods $|$ dislocation $|$ grain boundary $|$ materials design $|$ computational chemistry\newline}

\dates{This manuscript was compiled on \today}

\begin{abstract}

The fundamental construct of organic chemistry involves understanding molecular behavior through functional groups. Much of computational chemistry focuses on this very principle, but metallic materials are rarely analyzed using these techniques owing to the assumption that they are delocalized and do not possess inherent functionality. In this paper, we propose a methodology that recovers functional groups in metallic materials from an energy perspective. We characterize neighborhoods associated with functional groups in metals by observing the evolution of Bader energy of the central cluster as a function of cluster size. This approach can be used to conceptually decompose metallic structure into meaningful chemical neighborhoods allowing for localization of energy-dependent properties. The generalizability of this approach is assessed by determining neighborhoods for crystalline materials of different structure types, and significant structural defects such as grain boundaries and dislocations. In all cases, we observe that the neighborhood size may be universal---around 2-3 atomic diameters. In its practical sense, this approach opens the door to  the application of chemical concepts, e.g., orbital methods, to investigate a broad range of metallurgical phenomena, one neighborhood at a time.

\end{abstract}

\begin{document}

\maketitle
\thispagestyle{firststyle}

\abscontent

\section{Introduction}

Chemical properties such as reactivity in molecular systems originate from functional groups that are believed to retain their chemical identity regardless of the constitutive environment. Taking organic chemistry as the quintessential example, understanding the chemistry of few functional groups has proven extremely useful in predicting molecular behavior. The functional group concept has also been useful in studying and manipulating the properties of many classes of materials, like carbon nanotubes \cite{ago1999work}, metal organic frameworks \cite{collins2016materials}, and proteins \cite{arima2007effects}. Amongst those materials that have not benefited from this chemical analysis are metals---owing to the assumption that they are delocalized structures that do not possess an equivalent of functional groups. We plan to test this very assumption, by devising a methodology to determine and assess functionality associated with metallic materials.

Functionality is a well-studied chemical concept. An important early investigation by Bader and Beddall showed that it is possible to spatially partition molecules into volumes with well-defined energies and boundaries \cite{bader1972virial}. In his book, Bader mentions, ``The primary purpose in postulating the existence of atoms in molecules is a consequence of the observation that atoms or groupings of atoms appear to exhibit characteristic sets of properties (static, reactive, and spectroscopic), which, in general, vary between relatively narrow limits''.  He defines this grouping of atoms as a functional group. He further adds, ``In some series of molecules, the variation in the properties is so slight that a group additivity scheme for certain properties, including the energy, can be established''\cite{AIM}. When such a condition exists, these groupings of atoms are said to possess transferability - \textit{i.e.}, their energy and other properties are insensitive to the local environment. In his earlier work to assess transferability of a bonded fragment, Bader observed that the difference in charge distribution and energy of \ch{CO} in \ch{CO2} and \ch{OCS} is very small. He thus perceived a measure of transferability of functional groups being determined by the extent to which their charge distributions remain unchanged during transfer between systems \cite{bader2008nearsightedness, bader1971spatial, hologram}. These functional groups seemed to have an associated ``neighborhood'' wherein atoms felt its effect. Successive analyses that followed on the concept of transferability and its applications involved a rigorous investigation of charge distributions, and notably, not much attention was given to how the energy of functional groups evolve when moved from one constitutive environment to another \cite{sablon2010linear, fias2017chemical, bushmarinov2009atomic, gatti2013challenging, matta}.

A more recent perspective on functionality was presented by Prodan and Kohn as the theory of nearsightedness of electronic matter (NEM). NEM derives its name from the observation that atoms ``see clearly'' only nearby atoms \cite{NEM_PNAS, NEM_PRB}. Rigorously, it posits that for a fixed chemical potential, the charge density, $\rho(\bm{r})$, and local properties originating from $\rho(\bm{r})$ are sensitive to changes to the external potential within some neighborhood with nearsightedness range $R_c$. Changes to the potential beyond $R_c$---no matter how large---do not significantly effect the density. NEM has been argued to be the foundation of many chemical concepts including ``divide and conquer'',``chemical transferability'', and Pauling's ``chemical bond'' \cite{ Divide_conqer_anastassia, NEM_PRL, Divide_conqer2, Divide_conqer3, Pauling_1939, mcnaught1997compendium}.

To quantify NEM and transferability of functional groups, one must determine the neighborhood size wherein its effects are felt. In this context, of particular relevance is the work done in \cite{fias2017chemical} to determine the ``nearsightedness range''. The authors investigated changes to $\rho$ with changing molecular structure using the softness kernel. They argued that the effect of a functional group is felt only in a neighborhood about 3 atoms away, and structural changes that are farther from the neighborhood do not produce significant chemical effects. They showed that the correct view of nearsightedness and functionality is to picture it as a process under constant chemical potential consistent from the perspective that reactions are carried out in a solvent. 

However, there is another equally valid way to envision functionality---more in keeping with Bader's idea of additivity---as the successive addition of environments around a functional group. In accordance with this view, we consider change in energy associated with adding successive coordination spheres about a central atom, and investigate the viability of these energy changes to determine neighborhood size, consequently serving as a means to describe metallic functionality.  We propose a methodology to assess $\rho$ sensitivity in these systems through changes in Bader energy. Identifying these neighborhoods in metals enables localization of structural perturbations that significantly alter energy.

\begin{figure}
    \centering
    \includegraphics[width=0.5\linewidth]{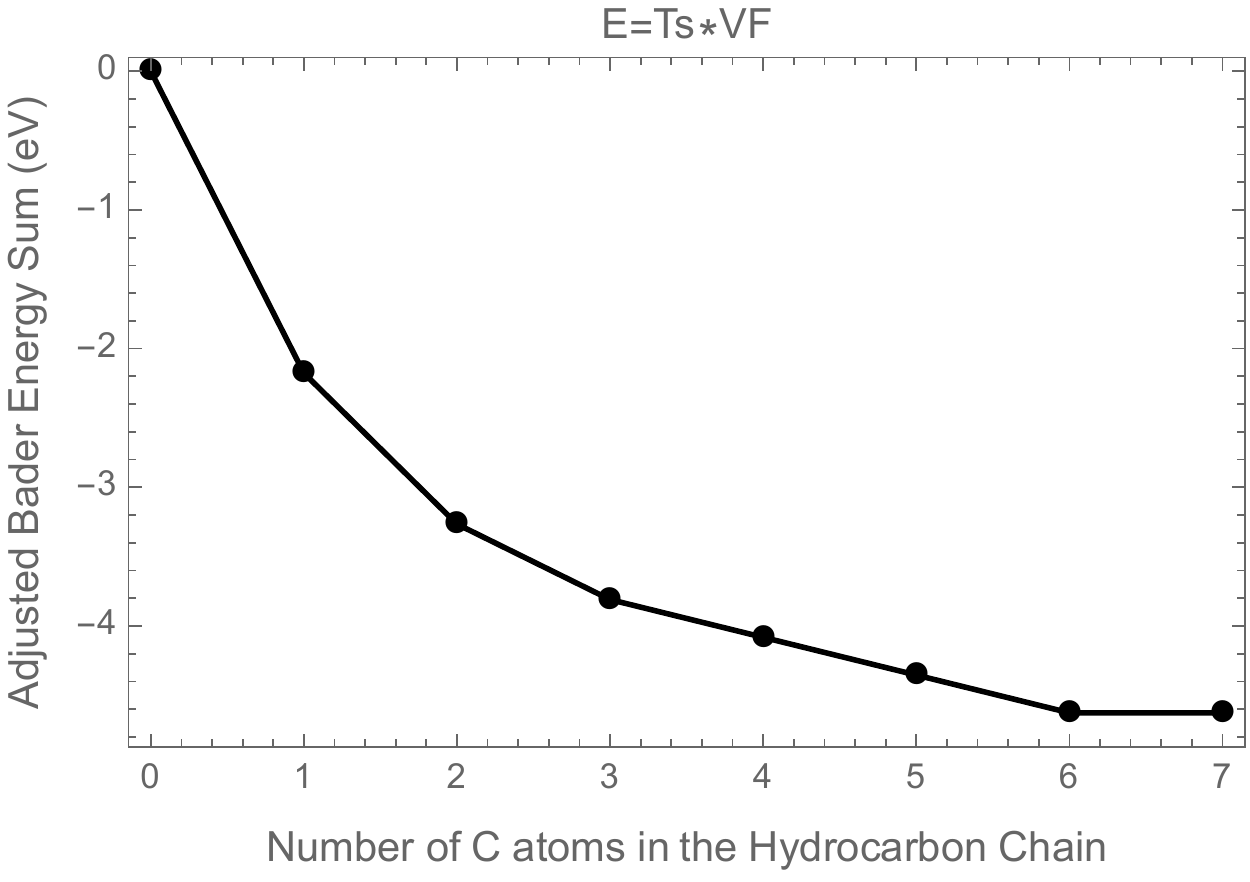}
    \caption{The evolution of energy of an aldehyde group with increasing carbon atoms in the attached hydrocarbon chain.}
    \label{fig:CHO}
\end{figure}

As an initial step in validating our approach, we explored the evolution of energy for a well-understood organic functional group, {\it i.e.}, an aldehyde, by growing the hydrocarbon chain attached to it. Our approach was aimed at quantifying the associated neighborhood of the aldehyde functional group. We considered \ch{CHO(CH_2)_nH} for our calculations, and varied n from 0 to 7. In accordance with the virial theorem, we utilized Bader partitioning to subsequently compute the energy of the aldehyde functional group (defined as the sum of the Bader energies of C, H  and O) for each step, as summarized in \cref{fig:CHO} \cite{AIM, matta}. 

Bader analysis of the energy produces many representations that account for the contributions from the exchange and correlation kinetic energy, including one being determined by the sum of the noninteracting and correlation kinetic energy ($T_s+T_c$), another being the product of the noninteracting kinetic energy and the virial factor ($T_s$*VF) \cite{Rodriguez_virial}. For our analysis we chose Bader energy to be represented by the latter ($T_s$*VF), as $T_c$ is very small in metals, and by using the virial factor we account for very small deviations from equilibrium around the central atom. Although all representations yield similar behavior, the energy obtained through this representation was observed to be most sensitive to structure changes near the functional group (see SI). The energy evolution thus determined is summarized in \cref{fig:CHO}.

Consistent with our expectations, the energy is an exponentially decaying function. The energy change beyond 3 carbon atoms from the functional group, is found to be less than three-tenths of an eV, coming to less than one-tenth of an eV between carbons 6 and 7. Our observations from energy evolution retrieve similar results from the softness kernel analysis in \cite{fias2017chemical}. Armed with these findings, we now move onto metals.

\section{Neighborhoods in Metallic Materials}

\begin{figure}
    \centering
    \includegraphics[width=0.5\linewidth]{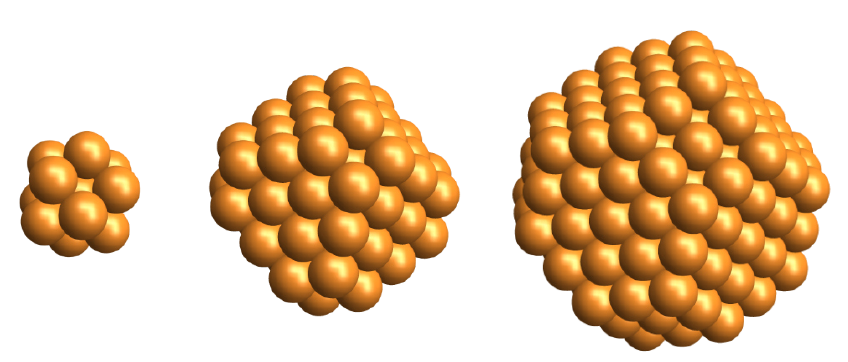}
    \caption{A schematic of the series of cluster calculations for metallic materials is presented here. The cluster to the left indicates the central cluster $\Omega$. The cluster in the middle indicates the central atom surrounded by 4 coordination spheres, and the one to the far right denotes a critical cluster with $n$ coordination spheres.}
    \label{fig:clusters}
\end{figure}

For the subsequent calculations presented in the paper, our general approach is to consider a ``central cluster'' $\Omega$ of radius $R_\Omega$ consisting of a central atom and its first coordination shell---{\it i.e.}, first nearest neighbors. We now isolate this cluster from its constitutive environment to compute the energy of the central cluster $\Omega$. We then compute the changes to this energy $\Delta \varepsilon_{\Omega}$ as the atomic environment at successively greater distances from $\Omega$ are restored. Simply, we add in concentric ``spheres'' containing second nearest neighbor atoms, third nearest neighbor atoms, and so on until $n^{\text{th}}$ nearest neighbor atoms. A schematic of this series of cluster calculations is presented in \cref{fig:clusters}. At some $R$ (or equivalently, $n$), we expect that $\Delta \varepsilon_{\Omega}$ approaches zero asymptotically. This critical cluster size, denoted as $R_c$, defines the neighborhood thus associated with the structure. Changes beyond $R_c$, no matter how large, do not significantly alter the energy of the central cluster.

Through the aforementioned methodology we were able to calculate the central atom energy $E^x_n$ as a function of adding successive coordination shells around it. Here $x$ indicates the element and $n$ indicates the number of coordination shells surrounding the central atom.  Adopting this notation, the energy of an isolated atom is represented as $E^x_0$ and the energy of single atom contained in an extended system as $E^x_{\infty}$. It then follows that the formation energy, $E_f$, of an elemental crystal  is given as $E_f = E^x_{\infty} - E^x_0$.  

When crystalline systems are analyzed as progressively larger clusters, there exists a decreasing perturbative influence on the central atom due to the increasingly distant free surface. This perturbation energy to the central atom as a function of $n$ is given by $\Delta E_n^x = \Delta E_n^x - \Delta E_0^x$. As $n$ keeps growing, the value of $\Delta E_n^x$ will approach the crystal formation energy $E_f^x$. This perturbative effect can also be equivalently interpreted as stabilization of the central atom (or central cluster) as a result of embedding in progressively larger coordination shells. Both interpretations offer conceptual advantages when seeking to understand metallic functionality.

We now define a stabilization function, \mbox{$\Delta \varepsilon^x_n \equiv \,\mid E^x_{\! \infty} - E^x_n \mid$}, giving the magnitude by which the energy of an atom in an extended system is perturbed by the complete removal of all material beyond its $n^{\text{th}}$ coordination shell.  Where the system of interest is a perfect crystal, this function may be placed in a convenient form by adding and subtracting $E^x_0$,

\begin{equation}
\label{equ:stabilization_function}
 \Delta \varepsilon^x_n \, = \,\mid (E^x_{\! \infty} - E^x_0) - (E^x_n - E^x_0) \mid  \, = \,\mid E^x_{\! f} - \Delta E^x_n \mid
\end{equation}
where $ \Delta E^x_n \equiv E^x_n - E^x_0  $ is a stabilization function giving the per atom energy change to $\Omega$ through the successive addition of $n$ coordination spheres.

\section{Computational Setup}

Eleven crystalline elements (Al, Si, V, Cu, Nb, Mo, Tc, Ru, Rh, Pd, Ag) were modeled as clusters created with $n$ varying from 0 to as large as 11. To model defect structures, we considered an Al dislocation, and grain boundaries in Cu and Fe which were also modeled as clusters. Density functional theory methods provided within the SCM chemistry and materials modeling suite \cite{ADF1, ADF3} were used. The per atom energy of $\Omega$ was found similarly using Bader partitioning.

ADF \footnote{amsterdam density functional} cluster methods were used to calculate $ \Delta E^x_n $, while the BAND \footnote{accurate periodic DFT code} package  was used to calculate $E_f$ \cite{ADF1, ADF3, band1, band2} for the crystals. \cref{equ:stabilization_function} was used to calculate $\Delta \varepsilon^x_n$ for a series of crystalline materials and defect structures, which we then use in a search for functionality. 

\section{Crystalline Neighborhoods}

For the crystal systems, we constructed our calculations in sets of increasing complexity. In the first set we determined $\Delta \varepsilon^x_n$  of four crystals: diamond cubic (DC) silicon, a prototype covalent material; face centered cubic (FCC) aluminum, a free electron metal; FCC copper, a  $d$-block metal with a full $d$-band; and body centered cubic (BCC) vanadium, a metal with a partially occupied $d$-band.  These elements were chosen both because of their different crystal and electronic structures, and also because they could be modeled using the same basis set for both the BAND and ADF cluster calculations, thus minimizing inherent error in the calculated values of $\Delta \varepsilon^x_n$ (see SI). The focus of the second set of calculations was to develop an understanding of the factors causing variation to the stabilization function across elements sharing the same crystallographic structure. For this set of calculations we drew  from the  4$d$ transition metals:  BCC  niobium and molybdenum;  hexagonal close packed (HCP) technetium and ruthenium; and  FCC rhodium, palladium, and silver. 

\subsection{Silicon, Aluminum, Vanadium and Copper}
\begin{table}[h!]
	\centering
	\caption{\label{table:DC}
		The diamond cubic (DC) shell structure.
		Row 1:~Number of the coordination shell.
		Coordination shell zero is the central atom.
		Row 2:~Number of atoms in coordination shell $\bm{n}$.
		Row 3:~Total number of atoms in the cluster of $\bm{n}$ coordination shells.
		(Hard sphere representations of some of these clusters are provided in the SI.)
		Row 4:~Radius of the cluster, {\it i.e.}~distance between the central atom and the atoms of the $\bm{n^{\text{th}}}$ shell in atomic diameters or equivalently nearest neighbor separations.}
	\begin{tabular}{|l|c|c|c|c|c|c|c|c|c|c|c|c|}
		\hline
		\textbf{Coordination shell $\bm{n}$}
		& \textbf{0}
		& \textbf{1}
		& \textbf{2}
		& \textbf{3}
		& \textbf{4}
		& \textbf{5}
		& \textbf{6}
		& \textbf{7}
		& \textbf{8}
		& \textbf{9}
		& \textbf{10}
		& \textbf{11}
		\\\hline
		Number of $n^{\rm{th}}$ neighbors
		& 0
		& 4
		& 12
		& 12
		& 6
		& 12
		& 24
		& 16
		& 12
		& 24
		& 12
		& 8
		\\\hline
		Total atoms in cluster
		& 1
		& 5
		& 17
		& 29
		& 35
		& 47
		& 71
		& 87
		& 99
		& 123
		& 135
		& 143
		\\\hline
		Cluster radius
		& 0
		& 1
		& $\text{2} \sqrt{\frac{\text{2}}{\text{3}}}$
		& $\sqrt{\frac{\text{11}}{\text{3}}}$
		& $\text{4}\sqrt{\frac{\text{1}}{\text{3}}}$
		& $\sqrt{\frac{\text{19}}{\text{3}}}$
		& $\text{2} \sqrt{\text{2}}$
		& 3
		& $\text{4} \sqrt{\frac{\text{2}}{\text{3}}}$
		& $\sqrt{\frac{\text{35}}{\text{3}}}$
		& $\sqrt{\frac{\text{43}}{\text{3}}}$
		& 4
		\\
		\hline
	\end{tabular}
\end{table}


\begin{table}[h!]
	\centering
	\caption{\label{table:Si-summary}
		Si atomic diameter, energy of formation, isolated atomic energy, and changes in central Bader atom energy resulting from the addition of cluster coordination spheres ($\bm{\Delta E^x_n}$) as described in the text.
		Distances are reported in \AA\ and energies in eV.
		$\bm{\Delta E_{10}}$ was not determined.
	}
	\begin{tabular}{|c|c|c|c|c|c|c|c|c|c|c|c|c|}
		\hline
		\textbf{Si diameter (\AA)}
		& $\bm{E_f}$
		& $\bm{E_0}$
		& $\bm{\Delta E_1}$
		& $\bm{\Delta E_2}$
		& $\bm{\Delta E_3}$
		& $\bm{\Delta E_4}$
		& $\bm{\Delta E_5}$
		& $\bm{\Delta E_6}$
		& $\bm{\Delta E_7}$
		& $\bm{\Delta E_8}$
		& $\bm{\Delta E_9}$
		& $\bm{\Delta E_{11}}$
		\\\hline
		2.352
		& -5.42
		& -7865.72
		& -3.05
		& -9.01
		& -5.66
		& -5.55
		& -6.09
		& -6.71
		& -6.67
		& -6.12
		& -5.91
		& -5.79
		\\
		\hline
	\end{tabular}
\end{table}

Silicon possesses the diamond cubic crystallographic structure. 
Its near neighbor shell structure is summarized in \cref{table:DC}, with the central cluster $\Omega$ represented by a five atom cluster (\ch{Si5}) composed of a single atom and its four tetrahedrally coordinated nearest neighbors. 

Relative to the energy of an isolated Si atom, $E_0$, the calculated per atom Bader energy of \ch{Si5} as boundary was increased to include a second, third, fourth and so on up to eleven coordination shells is reported in \cref{table:Si-summary} along with the nearest neighbor distance of Si and its calculated formation energy, $E_{\!f}$. 

\begin{figure*}[h]
\centering
\includegraphics [width=0.9\linewidth]{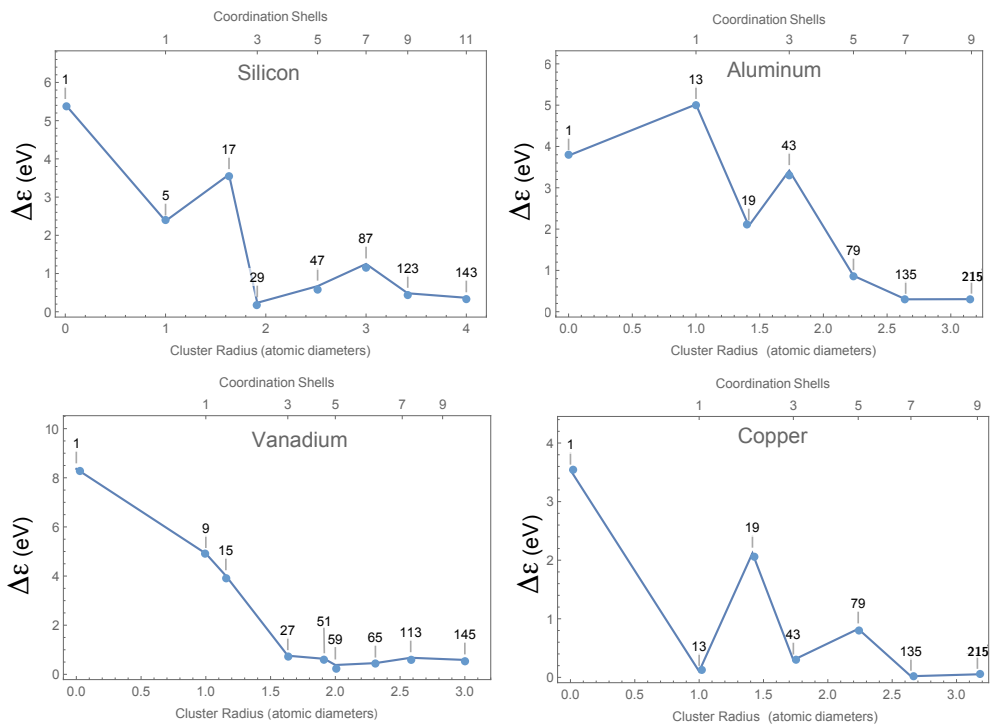}
\caption{\label{fig:groupI_energies}
	The stabilization function ($\Delta \varepsilon^x_n$) for the elements Al, Si, V and Cu as a function of $n$.
	The callouts in the graphs give the number of atoms in the representative cluster.
	For example, an Al cluster representing a central atom and its first 7 coordination spheres will contain 135 atoms.
}
\end{figure*}

These results are summarized graphically in the upper left of \cref{fig:groupI_energies} where $\Delta \varepsilon^{\text{Si}}$ is depicted as a function of $n$.
This figure represents the sensitivity of $\Omega$ to the retreating perturbation, that is, the distance over which the central cluster $\Omega$ can clearly see the free surface. When the energy goes to zero, from $\Omega$’s point of view, the surface has vanished. Plainly, when the free surface is infinitely distant from $\Omega$, the per atom energy of $\Omega$ will be identical to that of an atom in crystalline Si. Hence beyond some point the decay of $\Delta \varepsilon$ will approach zero asymptotically. In fact, because for both ordered and disordered gapped materials\footnote{The systems modeled here are all gapped, in that the central cluster energy converges to  within \textasciitilde 0.3 eV---our measure of significance---of the formation energy before the energy difference between the LUMO and HOMO is on the order of $k$T.} the change in the density due to a perturbation  at $R$ decays exponentially with $R$,\cite {NEM_PNAS} it is arguable that energy decay should not only be asymptotic but exponential. 

Regardless, inspection of \cref{fig:groupI_energies} reveals that the onset of the asymptotic/exponential decay begins with coordination shell seven, where the per atom energy difference between crystalline silicon and the central cluster $\Omega$ is on the order of an eV, decreasing to 0.36 eV at eleven coordination shells---thus defining the critical cluster size to be three atomic diameters beyond $\Omega$. 

Inherently, the variation of $\Delta \varepsilon$ originates from  changes to the central cluster charge density \cite{molecules_in_metals}.
And just as the energy of a central cluster within an infinitely distant free surface will be equivalent to that of a crystal, so too will its  charge density  be identical to the crystalline density. For all elemental crystals, equivalence between the central cluster and crystalline charge densities is required when the Bader atom surfaces of the central cluster are coincident with the crystalline Voronoi polyhedra (cells) about each atom. Quite generally, the difference between the surface of a cluster's Bader atoms and a crystal's Voronoi cells provides a measure of their charge density differences, which vanish when the two surfaces coincide \cite{hologram}. 

\begin{figure}
	\begin{center}
		\includegraphics[width=0.5\linewidth]{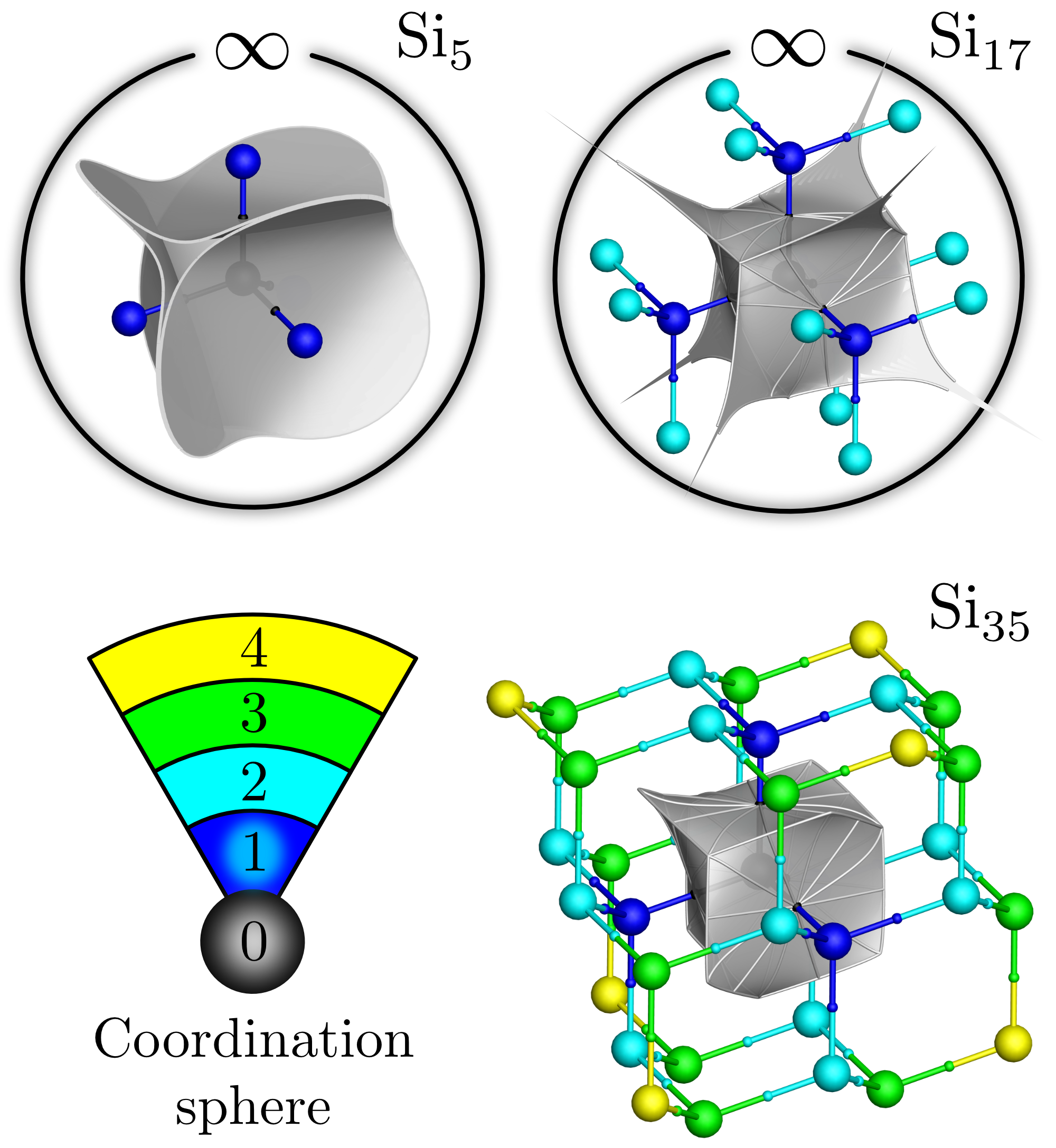}
	\end{center}
		\caption{\label{fig:Si_bader_atoms} Depiction of the central Bader atom surfaces in 5, 17, and 35 atom clusters (1, 2, and 4 coordination shells respectively).
		Nuclear positions are indicated by spheres colored according to coordination shell.
		In \ch{Si5} (top-left) the central Bader atom remains clearly open in four regions, one of which is facing to the left of center, while elsewhere the Bader atom surfaces are essentially converged.
		With two coordination spheres (top-right; \ch{Si17}) those same open regions have closed coincident with topological cage points, and only slivers of very-nearly converged surfaces prevent the Bader atom from being closed.
		At four coordination spheres (bottom-right; \ch{Si35}) the Bader atom has closed completely.
		The open surfaces in \ch{Si5} were truncated according to the $0.001\text{e}^-$ charge density isosurface.}
\end{figure}

A Bader atom's surface must contain local charge density minima. (In the chemical literature \cite{AIM} these minima are called cage critical points to indicate that there is one such point interior to cages of bound atoms.) 
Plainly, an atom's local charge density minima may be a finite or an infinite distance from the atomic nucleus. If all the local minima are a finite distance from the nucleus, the surface of the Bader atom is topologically connected and the atom is said to be closed. On the other hand, if even one local minimum is located at infinity, the Bader atom surface is disconnected and the atom is said to be open. Importantly, for any open Bader atom there is a path lying entirely within the atom that runs from the location of the nucleus to a point at infinity. Quite simply, an open atom is characterized by a channel of decreasing charge density extending from the nucleus to the neighborhood of at least one infinitely distant point. In contrast, the surfaces of crystalline Voronoi polyhedra are necessarily connected. For example, the Voronoi cell of the diamond cubic structure is in a class of truncated tetrahedra. 

As a means of clarifying this issue, consider the evolution of $\Omega$'s Bader atoms as more coordination shells are added \cite{molecules_in_metals}. In the case of Si, this process is represented in \cref{fig:Si_bader_atoms}. The top-left frame depicts the Bader atom surfaces of \ch{Si5}, or equivalently the interatomic boundaries between the central atom and its first coordination sphere. This set of surfaces is constructed from four asymptotic---hence disconnected---surfaces. As a result, all  the Bader atoms of the \ch{Si5} cluster are open. In other words, around every point there is a direction in which the charge density is decreasing and thus local minima are infinitely distant from the central atom. 

The boundary of the central Bader atom evolves with the addition of the twelve atom second coordination sphere to yield the \ch{Si17} cluster pictured in the top-right frame of \cref{fig:Si_bader_atoms}.  While the asymptotes separating the surfaces of this atom become steeper, the central Bader atom still extends to infinity. That is, running through all points near the central atom there is a path of decreasing charge density that leads to infinity. This path will pass through the ``spikes'' evident in the top-right frame of \cref{fig:Si_bader_atoms}. 

As depicted in the bottom-right frame of \cref{fig:Si_bader_atoms}, it is with the addition of the fourth coordination sphere to make a \ch{Si35} cluster that the boundary of the central Bader atom is topologically connected and the atom is closed. It is at this point that the cages of bound atoms sharing the central atom as a common vertex are completed, which mandates a single local minimum at the center of each of these cages. In a sense, it is with the completion of these cages, and the resultant closing of its Bader atom, that the central atom becomes isolated from the surroundings through an intervening shell of charge density. 
							
While the central atom is closed with the fourth coordination shell, $\Omega$ closes at the seventh coordination shell, {\it i.e.}~\ch{Si87}. It is here that the cages having a first coordination sphere atom as a vertex are completed. In fact, at this point the Bader atoms of $\Omega$ possess the topology of the Voronoi polyhedra of crystalline silicon. And, though \cref{fig:groupI_energies} shows only a few points beyond the seventh coordination shell, $\Delta \varepsilon$ decreases monotonically (arguably exponential decay) through these points. Closing of the Bader atoms of the $\Omega$ provides a means to identify and characterize these crystalline neighborhoods with well-defined energies.

Returning to the remaining crystals of the first set:~Al, V, and Cu. The central cluster of the FCC metals (Al and Cu) is a cuboctahderon consisting of a central atom and its twelve nearest neighbors. The central cluster of BCC V is a nine atom cube with an atom at the cube center and its eight nearest neighbors located at the cube vertices. Pictures of these clusters  along with analogues of \cref{table:DC} giving more information regarding the shell structure of FCC and BCC crystals along with the element specific cluster energy analogues of \cref{table:Si-summary} are provided in the SI.

This information is also summarized graphically in \cref{fig:groupI_energies} where $\Delta \varepsilon_n$ for the central cluster of each element is plotted as a function of $n$. In all cases, within ten coordination spheres $\Delta \varepsilon$ converges to within a fraction of an eV of the computed crystalline formation energy ($\Delta \varepsilon^{\text{Al}}_{9},\; \Delta \varepsilon^{\text{V}}_{10}, \; \Delta \varepsilon^{\text{Cu}}_{9} = $ 0.30, 0.58, 0.05 eV respectively). More noteworthy however, though the onset of exponential decay is element dependent---two, three, and five coordination spheres for V, Al, and Cu respectively---beyond the point where the Bader atoms of $\Omega$ close (BCC clusters of five coordinations spheres and 59 atoms, and FCC clusters also of five coordination spheres and 79 atoms) $\Delta \varepsilon$ is monotonically decreasing or level.

\subsection{4\textit{d} Metals}

\begin{figure*}[h]
\centering
\includegraphics [width=1.0\linewidth]{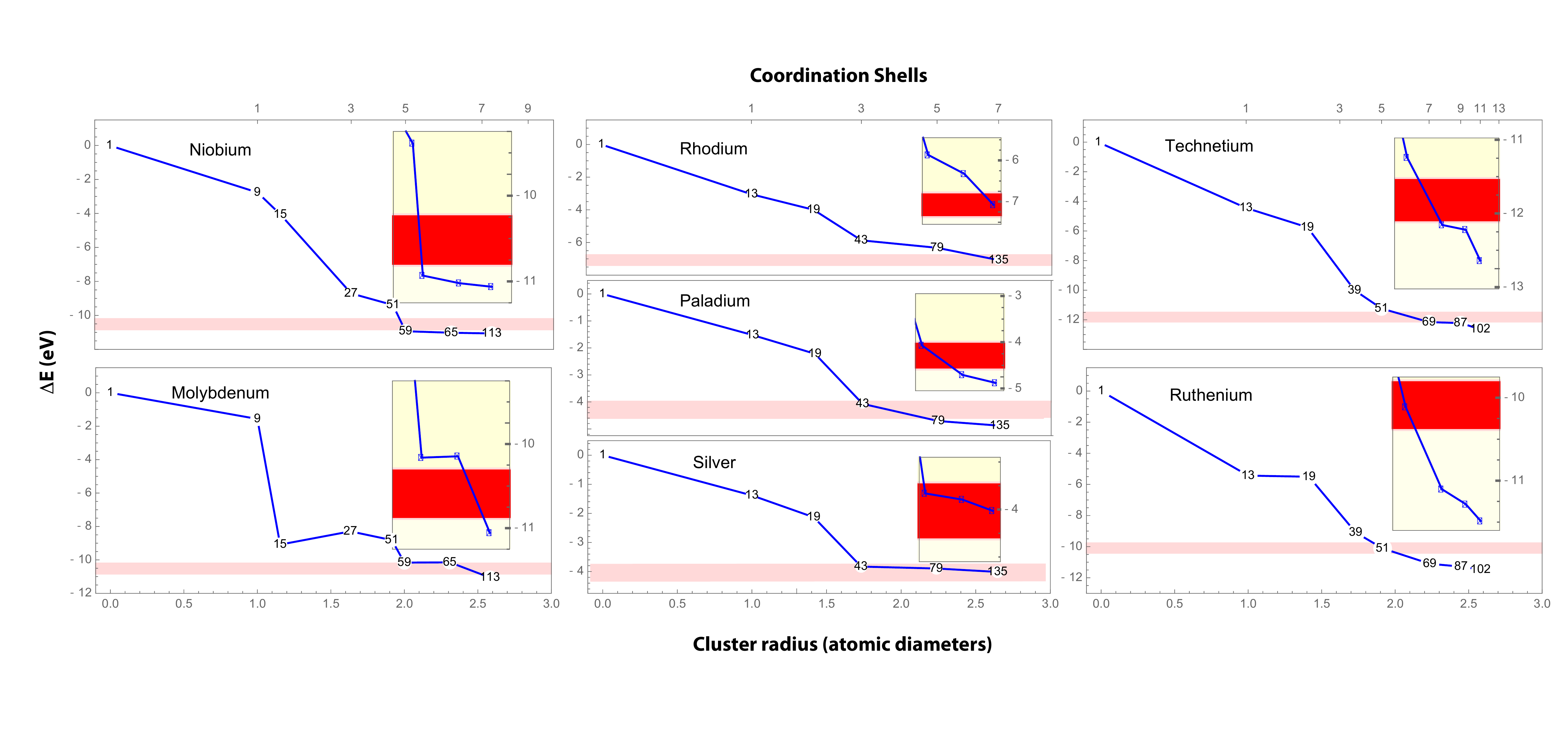}
\caption{\label{fig:group2_energies}
	The change in central Bader atom energy resulting from additional cluster coordination shells ($\Delta E$) for the BCC (left column), FCC (center column) and HCP (right column) 4$d$ transition metals.
	The red stripes indicate a range for the crystalline formation energy of $\pm0.3$ eV arising from different basis sets.
	Insets provide a higher fidelity depiction of $\Delta E$ for the three largest clusters of the series.}

\end{figure*}

 The 4$d$ transition metals required larger basis sets for the accurate determination of Bader atom energies. Because these basis sets were not available to the BAND calculations, a ``basis set mismatch'' error was introduced into the determination of $\Delta \varepsilon$.   Specifically, we estimate that the formation energy calculated using the smaller basis sets available to BAND may be as much as 0.6 eV less than its estimated value  from cluster calculations using larger basis sets (see SI).  Nonetheless, for this set of calculations we are more interested in the stabilization function.  Accordingly, $\Delta E$ as a function of the number of coordination shells for the BCC metals Nb and Mo; the FCC metals Rh, Pd and Ag; and the HCP metals Tc and Rh are shown in \cref{fig:group2_energies}. The information contained in these figures is provided in tabular form in the SI.


Consider as a first case the  similarities in the form of $\Delta E$ for the BCC metals Nb and Mo. Recall that the central nine atom cluster of a BCC metal closes at 59 atoms or equivalently five coordination spheres and 2.0 atomic diameters. \Cref{fig:group2_energies} reveals a rapid decrease of more than an eV in the central cluster energy on addition of the fifth coordination sphere---transforming a 51 to a 59 atom cluster;
and for both Nb and Mo bringing the central cluster energy to within 0.3 eV of the estimated formation energy. While in V (\cref{fig:groupI_energies}) the effect is not as dramatic, a similar central cluster stabilization is observed. This substantive change in $\Delta E$ is indicative of a strong coupling between the central nine atom cluster and the region extending from the fifth coordination sphere.

In addition to these similarities there are notable differences, most striking is the stabilization of the central cluster $\Omega$ by the third and fourth coordination shells. For Mo there is a dramatic stabilization of $\Omega$ with the addition of the third coordination shell, while for Nb  and V the same stabilization results  from the addition of the third and fourth coordination shells. These variations reflect  stronger coupling between $\Omega$ and its immediate surroundings in Mo compared to  Nb and V.  A relevant consequence for such a variation is that  in Mo the effects of a perturbation altering the positions of a few atoms will be more localized than those from the same perturbation in Nb or V.

The decay of $\Delta \varepsilon$ for the FCC metals---shown in the center column of \cref{fig:group2_energies}---also exhibit similar forms. Notably, the closing of $\Omega$ at five coordination spheres and 79 atoms is well inside the region of asymptotic decay for all the FCC metals modeled (Al, Cu, Rh, Pd, Ag).  And it is at this size that the energy of $\Omega$ falls to within 0.15 eV of the estimated formation energy for all the 4$d$ FCC transition metals. However, more striking is the central cluster stabilization from the third coordination shell, which for the 4$d$ transition metals accounts for the bulk of the stabilization energy.  This stabilization can be attributed to the partial closing of $\Omega$. 

Unlike the BCC metals, characterized by a single type of local minimum, the FCC metals have two: the first at the center of the FCC octahedral hole, and the second at the center of the FCC tetrahedral hole. The tetrahedral holes of the central atom form with the addition of the first coordination sphere and the central atom octahedral holes with the second coordination sphere.  The addition of the third coordination sphere at a cluster size of 43 atoms completes all the tetrahedral holes of $\Omega$.  And obviously the octahedral holes of $\Omega$ are completed at the same time as the closing of $\Omega$ with the addition of the  fifth coordination shell.  Even so, unlike Cu, for the 4$d$ FCC metals, there is little coupling between the central cluster  and the fifth coordination shell.  Hence, just as in the case of the BCC metals, the energy of a perturbation to a small number of atoms in the FCC transition metals will be differentially distributed in accordance with metal's stabilization function.  Particularly relevant to our subsequent discussion, perturbation energy will be more delocalized in Cu compared to the 4$d$ FCC metals.   

Lastly, turning to the modeled HCP transition metals where there are four symmetry unique local minima adjacent to $\Omega$.  Like the FCC structure there are both tetrahedral and octahedral minima. And like the FCC metals, the onset of exponential decay   begins with the addition of the second coordination sphere. However, with $\Omega$ possessing D$_{3h}$ symmetry, the octahedral and tetrahedral holes  are split into symmetry unique pairs depending on their displacement perpendicular or parallel to the 3-fold axis. The central thirteen atom cluster undergoes significant closure with the third coordination shell and completely closes with the fifth coordination shell---a cluster of 51 atoms.  And beyond which, the stabilization energy due to successive coordination spheres is a fraction of an eV.

The key point is that across  all structure types modeled,  the energy of $\Omega$ ``improves'' in lock step with the progressive closing  of $\Omega$ and falls to values that are comparable to thermal energies  when the cluster is fully closed by a surrounding volume  with a radius of 2.5 to 3 atomic diameters. In other words, this indicates that the size of the neighborhoods for all types of crystals, regardless of the type of element or structure, is consistent. The precise form of the stabilization function depends on the  arrangement of local charge density minima within this volume. 

\section{Defect Neighborhoods}

Defects play a significant role in dictating the properties and performance of materials. 
For example, the theoretical strength of mild steel, correlated with its Young's Modulus, ranges anywhere from 10-40 GPa \cite{dieter1976mechanical}. However, its measured tensile strength is much lower - between 0.7-2.1 GPa. This discrepancy between theoretical strength and measured strength is attributed to structural defects in the material.

Amongst the more important structural defects affecting performance in metals are dislocations and grain boundaries. A grain boundary is the region of transition between two atomic arrangements and maybe manipulated to enhance strength. Dislocations are characterized by a partially missing plane of atoms. Their production and mobility, which is the source of plastic deformation, also influence strength. Inhibiting their motion can lead to undesirable properties such as brittle failure. This is sufficient evidence to believe that these defect structures inherently possess functionality, and to determine defect neighborhoods wherein such energy dependent properties can be localized, we now turn to investigating the central cluster energy evolution in these systems.

\subsection{Al Edge Dislocation}

\begin{figure}[h]
	\begin{center}
		\includegraphics [width=1.0\linewidth]{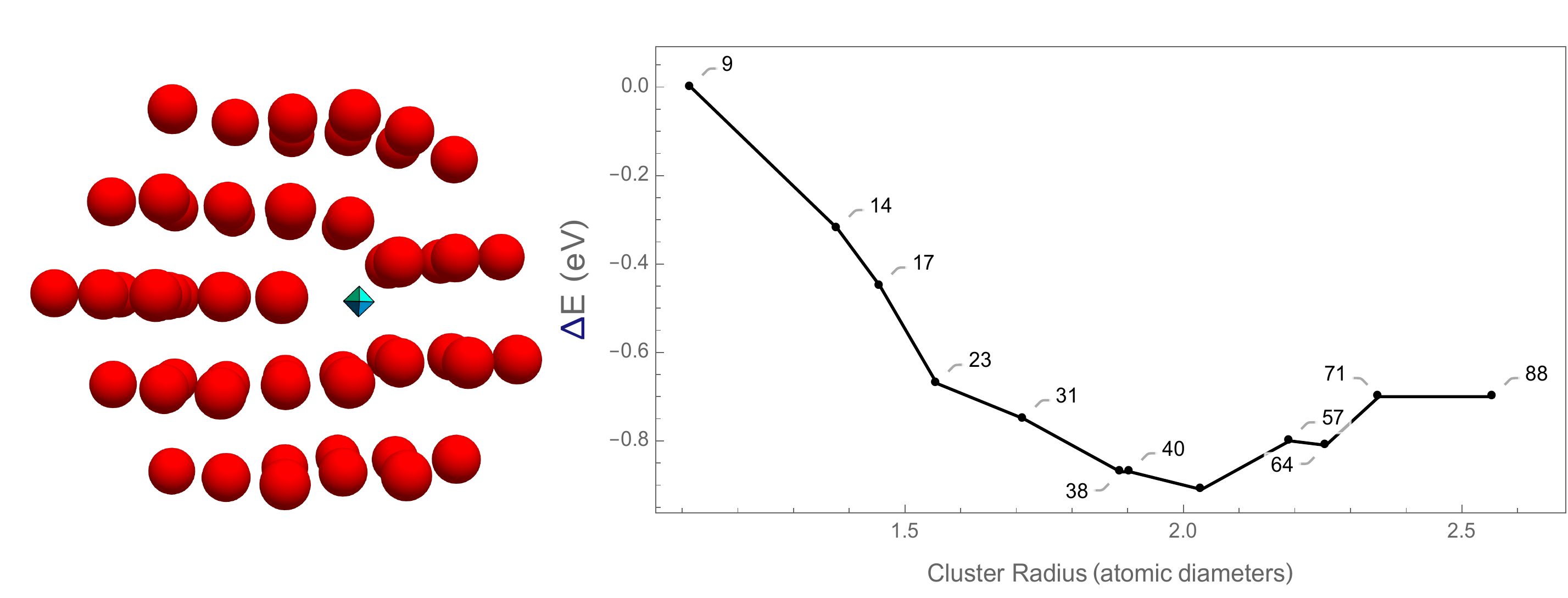}
	\end{center}
		\caption{\label{fig:Al_dislocation}
		Schematic representation of Al edge dislocation critical cluster - the red spheres indicate atoms and the cyan octahedron indicates the cage point on which the cluster is centered (left). Change in Bader energy per atom ($\Delta$E) for hole-centered $\Omega$ resulting from additional dislocation cluster coordination shells. Callouts indicate number of atoms constituting the representative clusters. (right)}
\end{figure}

An Al edge dislocation was generated via molecular dynamics (see SI). $\Omega$ for this structure was defined to be the 9-atom cluster centered on the non-crystallographic hole intrinsic to the defect. As in the crystalline studies, we computed the per atom stabilization function ($\Delta E$) for this cluster through the addition of successive coordination shells (though a coordination shell is somewhat arbitrarily defined for a non-crystalline system) about the central cluster $\Omega$. The calculated energies and specific structure geometries are summarized in the SI and the results are shown graphically in \cref{fig:Al_dislocation}. 

We observe that $\Omega$ is closed by the coordination shell that is 57 atoms and about 2.2 atomic diameters distant from the cluster center. At this point, the adjoining cages that have one of the atoms in $\Omega$ as a vertex are complete. Beyond this critical cluster size, the central cluster energy increases and then, as in the crystalline case, appears to asymptotically approach $E_{\! \infty}$, which denotes the energy of $\Omega$ when contained in an extended system.

Not unlike the crystalline calculations, the central cluster of the dislocation undergoes stabilization as we grow its surrounding coordination shells. From \cref{fig:Al_dislocation}, it is clear that the central cluster $\Omega$ undergoes a stabilizing effect up until it is closed---at which point subsequent addition of atoms destabilize $\Omega$. This is visibly evident from the increase in $\Delta E$ beyond 2.0 atomic diameters. As the system as a whole must adopt an electronic distribution of lowest energy, this destabilizing effect is permitted only if there is an offsetting energy lowering of the atoms making up the shells beyond 2.0 atomic diameters. 

\subsection{Cu and Fe Grain Boundaries} 

\begin{figure}[h]
	\begin{center}
		\includegraphics [width=0.6\linewidth]{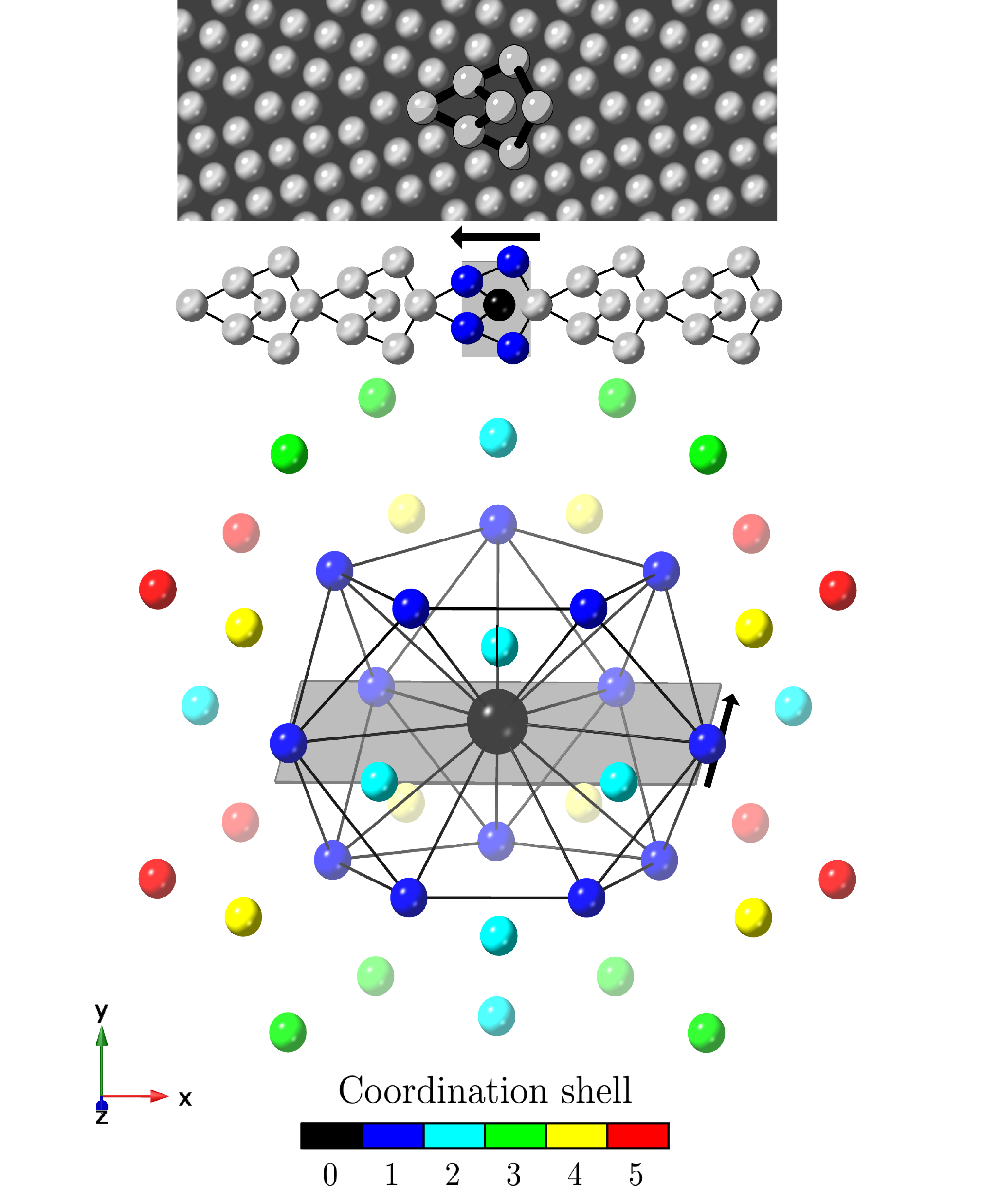}
	\end{center}

		\caption{\label{fig:Cu} Depiction of the symmetric 36.8$^\circ$ (310) <001> type tilt boundary cluster. A simulated image of the grain boundary region from reference \cite{grainboundary} with the structural unit is indicated (top). The 2-dimensional repeating structural unit (middle), and the corresponding 3-dimensional central cluster with the Cu atom in the center (bottom) are shown. The atoms are color coded based on their positions in the coordination sphere; and the central cluster is represented by the first coordination sphere. The XZ plane, indicated by the gray plane in the bottom frame coincides with the grain boundary plane and corresponds to the grey shaded rectangle in the middle frame, where the same atom coloring is used to show the central cluster atoms in the boundary. }

\end{figure}

A well-studied grain boundary both experimentally and computationally is a symmetric 36.8$^\circ$ <001> (310) type tilt boundary in Cu which Duscher {\it et al.},\cite{grainboundary} determined via atomic resolution x-ray diffraction to be characterized by a repeating kite structure represented schematically in \cref{fig:Cu}. The atoms immediately adjacent to this site form a roughly pentagonal coordination shell of 14 atoms.  We take this 14 atom shell containing a central Cu atom as the central cluster $\Omega$ for our investigation.  For this symmetric grain boundary coordination, this cluster is dominated by tetrahedral holes arranged in a manner incommensurate with crystal packing \cite{ashby}.


\begin{figure}[h]
	\begin{center}
		\includegraphics [width=0.7\linewidth]{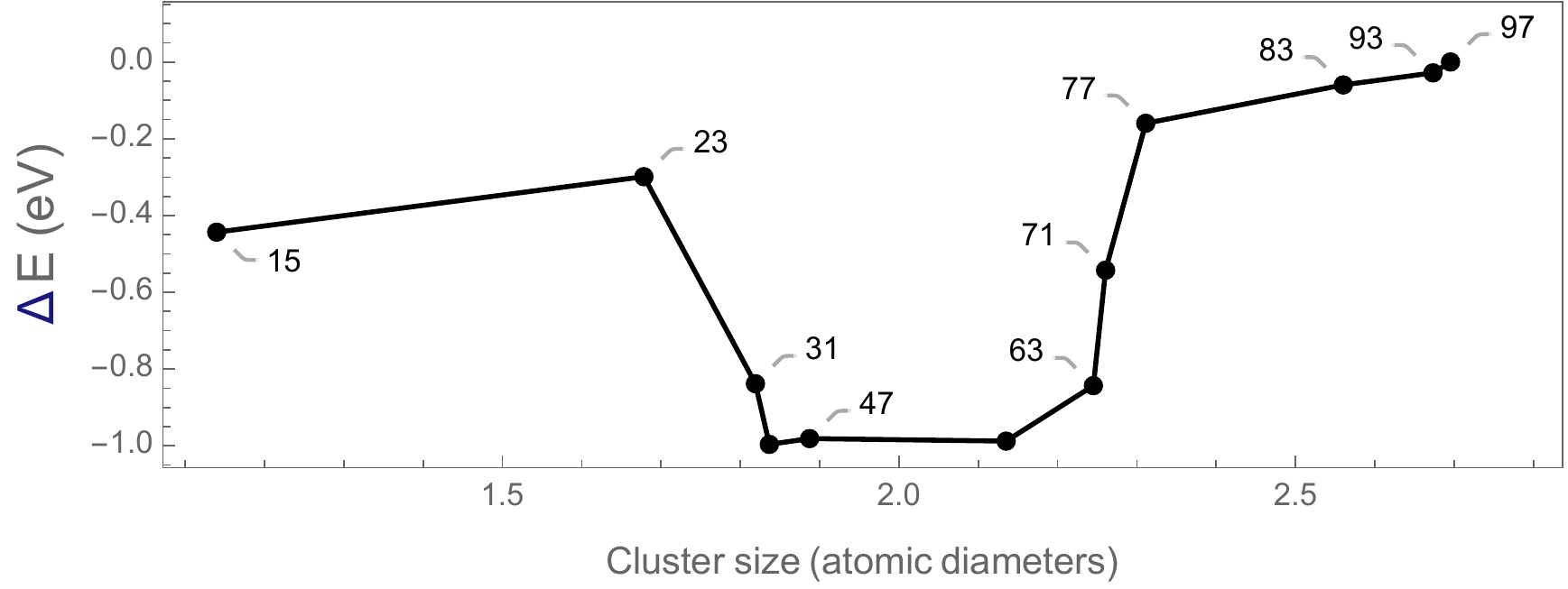}
	\end{center}
		\caption{\label{fig:Cu_energies}
		Change in central cluster Bader energy resulting from additional grain boundary cluster coordination shells.
		\textbf Per atom $\Delta E$ for Cu atom centered grain boundary clusters. Callouts indicate number of atoms constituting the representative clusters.}
\end{figure}

We similarly computed the stabilization function for the central cluster through larger coordinations. The calculated energies are given in the SI and the results are shown graphically in  \cref{fig:Cu_energies}. The grain boundary central cluster is fully closed by the coordination shell 2.3 atomic diameters distant from its center and containing 77 atoms. And as with the crystalline systems, beyond this point the stabilization function for Cu containing clusters appears to be asymptotically converging on $E_{\! \infty}$ which we estimate to be at 2.7 atomic diameters and 97 atoms (see SI)\footnote {using the number of points in \cref{fig:Cu_energies}, we could more accurately determine $E_{\! \infty}$ for the Cu grain boundary than we did for the Al dislocation, thus confirming that $\Delta E$ for these defects ultimately converges to $E_{\! \infty}$}. 


As in the case of the Al dislocation, inspection of \cref{fig:Cu_energies} reveals two regions of significant coupling to the Cu containing central cluster. The first region, which again clearly is stabilizing, occurs with the shells between 1.7 atomic diameters containing 23 atoms and 1.9 atomic diameters containing 47 atoms, and correlates to the closing of the non-crystallographically arranged tetrahedral holes adjoining  the central cluster. 

The second region that destabilizes the central cluster, occurs with the shells between 2.2 atomic diameters containing 63 atoms and 2.3 atomic diameters containing 77 atoms. Again, this is only possible through an offsetting energy lowering of the atoms making up the shells between 2.2 and 2.3 atomic diameters.

\begin{figure}[h]
	\begin{center}
		\includegraphics [width=0.7\linewidth]{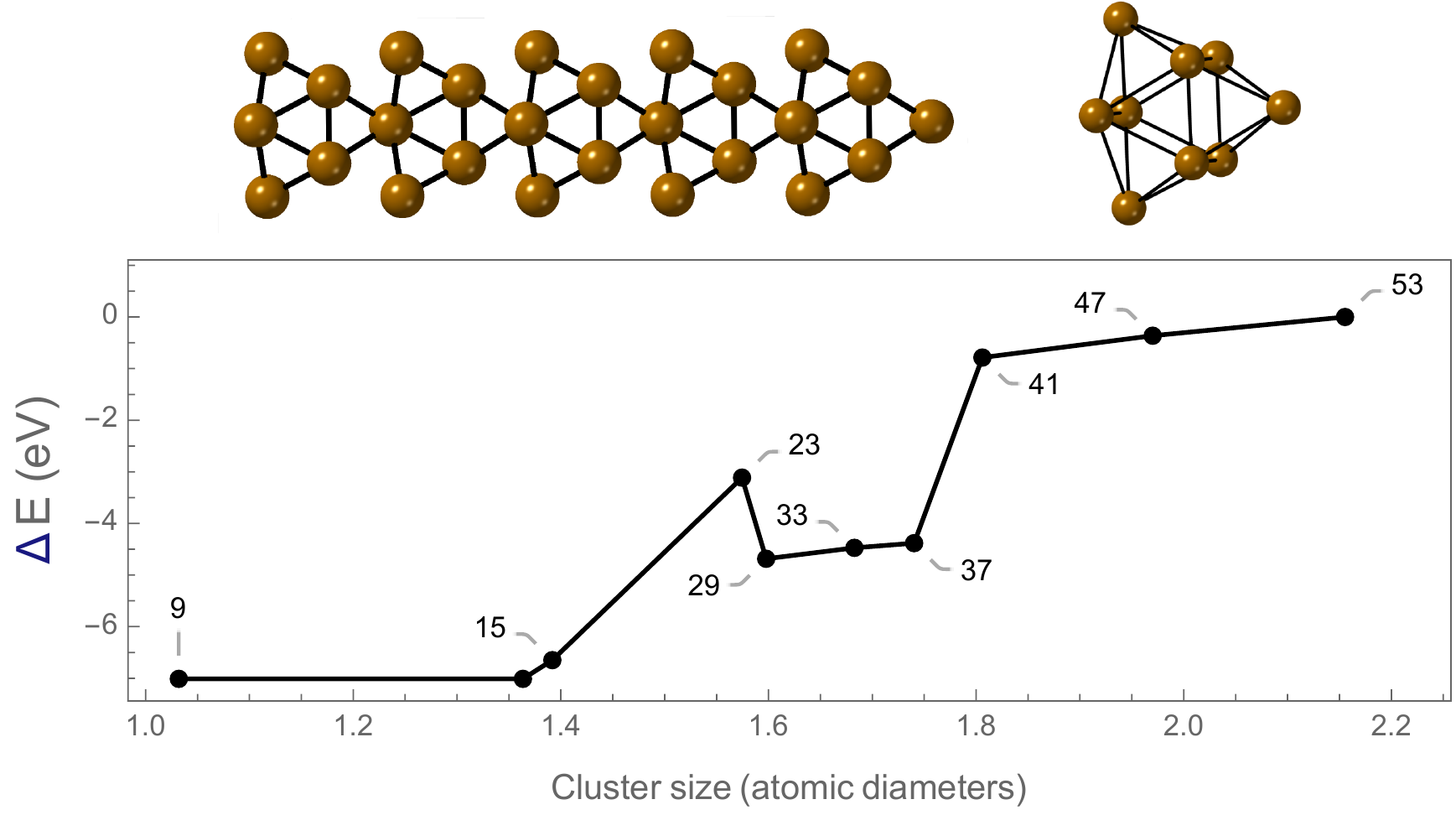}
	\end{center}
		\caption{\label{fig:Fe_energies}
		Schematic representation of the grain boundary and its 9-atom central cluster (top). Change in central cluster Bader energy resulting from additional grain boundary cluster coordination shells for FCC Fe.
		$\Delta E$ for Fe cage-centered grain boundary clusters. Callouts indicate number of atoms constituting the representative clusters.}
\end{figure}

To model another defect neighborhood, we looked at a symmetric 53.1$^\circ$ (210) <001> type tilt boundary in FCC Fe. We chose the central cluster $\Omega$ this time encapsulating a grain boundary cage point, defined by a 9-atom trigonal capped prism (\cref{fig:Fe_energies}). The stabilization function for $\Omega$ also reveals a similar trend to the Cu grain boundary calculations. At 1.6 atomic diameters and 29 atoms, the non-crystallographic tetrahedra adjoining the trigonal capped prism are closed, leading to the characteristic stabilization visible by a decrease in $\Delta E$. $\Omega$ in this case is closed at 1.8 atomic diameters and 41 atoms---when the tetrahedral and octahedral cages immediately adjoining $\Omega$ are closed. 

\section{Discussion}

We conjecture that the stabilization of the local environment in the case of metal or defect clusters are primarily due to the  presence of two-body non-directional $\sigma$ bonding. As in the case of defect calculations, we observe an additional destabilization as we incorporate more of the surrounding crystallographic region by growing the cluster. This effect can be explained by the presence of $\pi$ or $\delta$ bonds that have a strong angular dependence. In order to stabilize the distant crystallographic environment, the local structure undergoes highly directional $\pi$ or $\delta$ overlap that competes with the $\sigma$ overlap, leading to the observed increase in $\Delta E$.

In search of support for this conjecture, we turn to metal clusters. There is much experimental and theoretical evidence that the structure of small metallic clusters are far from crystallographic \cite{baletto2019structural, cohen1987physics, clemenger1991spherical, de1987electronic, montano1989structure, wertheim1989electronic}. For example, experimental characterization to determine the crystal structure for Ag and Fe microclusters showed that clusters that have radii less than 3 atomic diameters are inconsistent with crystal packing \cite{wertheim1989electronic}.

Our results provide an important insight to behavior of metals---there is a competition between the charge density produced by the atoms in the first few layers surrounding the central cluster, and the charge density produced by the rest of the layers in the neighborhood at large cluster sizes. Some structures may prefer a charge distribution that lowers local energy, while others promote charge distributions that lower system energy, {\it i.e.}~crystalline charge densities. Which of these two charge densities predominates is mediated by the respective stabilization functions. Regardless of the mechanisms responsible for the decay of $\Delta E$, any associated structural or chemical perturbation will produce a significant response in their respective neighborhoods.

We are able to recover neighborhoods associated with crystals, dislocations and grain boundaries that constitute functional groups in metallic materials---allowing for localization of energy-dependent properties. From the calculations it is evident that regardless of their shell structure and across various structure types, metallic functional groups appear to have a universally similar neighborhood size---they appear to be converging asymptotically (on a value near their estimated formation energies) within a radius of 2 to 3 atomic diameters. Equally as interesting, however, is the variation in the decay of $\Delta E$ across a given structure type. 

Depending on the structure and underlying chemistry, these systems can either promote local or crystalline bonding, which in turn affects the mechanical properties of the material. For example, some impurities at metallic grain boundaries are embrittling whilst others have been observe to enhance cohesion \cite{gibson2015segregation}. This difference in behavior originates from changes to atomic interaction between neighborhoods. These neighborhoods thus serve as adequate models for further and more chemically based investigations  to uncover the relationships between electron density and the mechanical response of metallic materials.

\section*{Acknowledgments}

The authors would like to acknowledge the contribution of Dr.\ Garritt Tucker and Jacob Tavenner from the Computational Materials Science and Design group at the Colorado School of Mines for providing  grain boundary and dislocation structures.
We also acknowledge valuable discussions with Dr.\ Travis Jones from the Fritz Haber Institute of the Max Planck Society.

\section*{Funding}
Support of this work under ONR Grant No.~N00014-10-1-0838 is gratefully acknowledged. 

\section*{Data Availability}
The raw and processed data required to reproduce these findings are accessible from the supplementary information.

\bibliography{Refs}

\newpage



\section*{Supplementary material}

# Neighborhoods and Functionality in Metals (Supplementary Information)

M. Rajivmoorthy    T. R. Wilson    M. E. Eberhart

October 20, 2021

# Contents

# 1  Comments on Computational Approach

Central to this investigation is the comparison of Bader atom energies with energies obtained with band methods. While conceptually undemanding, in practice these comparisons introduces sources of computational error. Band determined energies exploit variational methods while Bader atomic energies are calculated via the virial theorem [1, 2, 3]. In the latter case, even small errors in the calculated core electron energies can produce large total energy errors, which may be minimized using all electron methods and large basis sets. In turn, to facilitate comparison of Bader atom and band energies, to the extent possible, the same all electron basis sets and computational framework should be used for both calculations. The Amsterdam Modeling Suite provides the capabilities necessary to address many of these issues.

The suite utilizes the Amsterdam Density Functional (ADF) package [4, 5] to calculate the electronic structure of clusters, and BAND [6, 7, 5] to model extended crystalline systems. Both codes use the same Slater-type-orbital (STO) basis functions, though some of the basis functions

available to ADF are not fully supported by BAND, most consequential, the all electron quadruple zeta basis set including four polarization terms (QZ4P).

This became a factor when modeling the $4d$ transition metals where a QZ4P basis set was required to calculate Bader atom energies accurately. Presumably the large basis set was needed due to the greater number of radial nodes and hence more rapidly varying near nucleus charge density. Nonetheless, since BAND calculations could not be converged using the QZ4P basis set a source of computational error was introduced when comparing BAND determined formation energies with Bader atom energies. As a way of estimating the magnitude of this error, ADF was used to calculate single atom and large cluster total energies for all the $4d$ transition metal elements using both the triple zeta including two polarization terms (TZ2P) and QZ4P basis sets. In general, and not surprisingly, the single atom energies were lower for the larger basis set by about 10 eV. For the larger clusters, the total energy per atom was again lower using the QZ4P basis, but this time by about 10.6 eV per atom. Using the difference between the single atom and large cluster total energies as an approximation to the formation energy, the QZ4P basis set yields a more negative formation energy of approximately 0.6 eV per atom.

As a check on the accuracy of Bader atom energies we used the fact that over a Bader atom the integral of $\nabla^2\rho(r)$ should be identically zero [8, 9]. Deviations greater than $10^{-2}$ are deemed marginal and indicative of numerical error. We found that the integrated Laplacian of the charge density over the central Bader atom was sensitive to computational parameters. Best results were achieved with a high density Voronoi integration scheme (accint = 6), "very good" density fitting and an appropriate choice of basis set. Even so, in some circumstances the integral of $\nabla^2\rho(r)$ over the central Bader atom was slightly greater than $10^{-2}$. However, at all times the same value over the central atom and the atoms of its first coordination shell were within acceptable limits. It is for this reason that we monitored the per atom energy of a "central cluster"—the central atom and its nearest neighbor coordination shell—as a function of changing boundary width. All calculations employed the generalized gradient approximation using the Perdew–Burke–Ernzerhof exchange-correlation functional (GGA PBE) [10]. Finally, the BAND calculations used a quadratic tetrahedron method for numerical integration over the Brillouin zone, sampling a minimum of 16 $k$ points in the irreducible wedge.

# 2 Determining the Bader Energy

There are many representations of Bader energy from the ADF calculations run for QTAIM analysis. Two particular ones are the following: the sum of the noninteracting and correlation kinetic energy ($T_s + T_c$), another being the product of the noninteracting kinetic energy and the virial factor ($T_s$*VF) as determined by [3]. The Bader energy determined by the two quantities are summarized.

Table 1: Bader energy representations for the aldehyde. The first row indicates the number of carbon atoms that are in the attached hydrocarbon chain; the second row represents the sum of $T_s + T_c$ for C, H, and O; and the third row represents the sum of $T_s$*VF for C, H, and O in the corresponding molecules.

| $C_n$ | 0 | 1 | 2 | 3 | 4 | 5 | 6 | 7 |
|---|---|---|---|---|---|---|---|---|
| $T_s + T_c$ (eV) | -3110.26 | -3113.26 | -3113.80 | -3113.80 | -3113.80 | -3114.07 | -3114.07 | -3114.07 |
| $T_s$*VF (eV) | -3097.20 | -3099.38 | -3100.47 | -3101.01 | -3101.28 | -3101.56 | -3101.83 | -3101.83 |

# 3 Lattice data

The shell structure for the four lattice types (DC, BCC, HCP, and FCC) discussed in the main text is provided in Tables 2-5 and pictures of representative clusters shown in Figures 1-4.

Table 2: Face-centered cubic (FCC) shell structure. Row 1: Number of the coordination shell. Coordination shell zero is the central atom. Row 2: Number of atoms in coordination shell $n$. Row 3: Total number of atoms in the cluster of $n$ coordination shells. (Hard sphere representations of some of these clusters are provided in the SI.) Row 4: Radius of the cluster, i.e. distance between the central atom and the atoms of the $n^{\text{th}}$ shell in atomic diameters or equivalently nearest neighbor separations.

| **Coordination shell** $n$ | 0 | 1 | 2 | 3 | 4 | 5 | 6 | 7 | 8 | 9 | 10 |
|---|---|---|---|---|---|---|---|---|---|---|---|
| Number of $n^{\text{th}}$ neighbors | 1 | 12 | 6 | 24 | 12 | 24 | 8 | 48 | 6 | 48 | 24 |
| Total atoms in cluster | 1 | 13 | 19 | 43 | 55 | 79 | 87 | 135 | 141 | 189 | 213 |
| Cluster radius | 0 | 1 | $\sqrt{2}$ | $\sqrt{3}$ | 2 | $\sqrt{5}$ | $\sqrt{6}$ | $\sqrt{7}$ | $2\sqrt{2}$ | 3 | $\sqrt{10}$ |

Table 3: Body-centered cubic (BCC) shell structure. Table layout is identical to that of Table 2.

| **Coordination shell** $n$ | 0 | 1 | 2 | 3 | 4 | 5 | 6 | 7 | 8 | 9 | 10 |
|---|---|---|---|---|---|---|---|---|---|---|---|
| Number of $n^{\text{th}}$ neighbors | 1 | 8 | 6 | 12 | 24 | 8 | 6 | 24 | 24 | 24 | 8 |
| Total atoms in cluster | 1 | 9 | 15 | 27 | 51 | 59 | 65 | 89 | 113 | 137 | 145 |
| Cluster radius | 0 | 1 | $2\sqrt{\frac{1}{3}}$ | $2\sqrt{\frac{2}{3}}$ | $\sqrt{\frac{11}{3}}$ | 2 | $4\sqrt{\frac{1}{3}}$ | $\sqrt{\frac{19}{3}}$ | $2\sqrt{\frac{5}{3}}$ | $2\sqrt{2}$ | 3 |

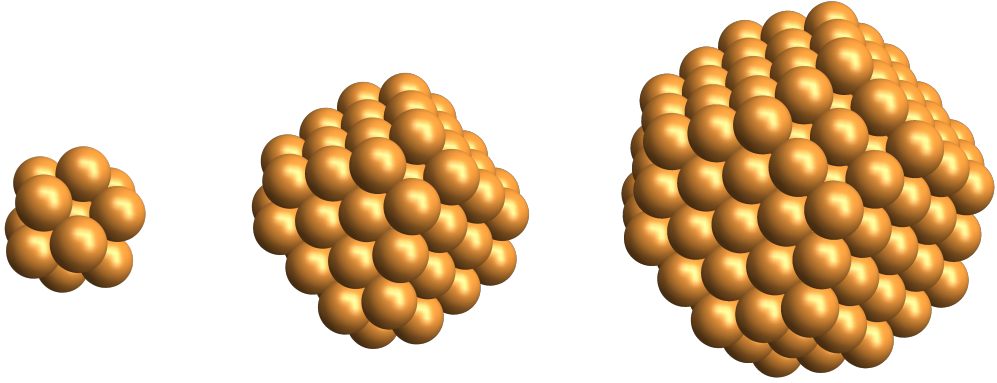

Figure 1: The FCC central 13-atom cluster (left), critical cluster consisting of 5 coordination shells (middle), and the largest cluster consisting of 10 coordination shells (right).

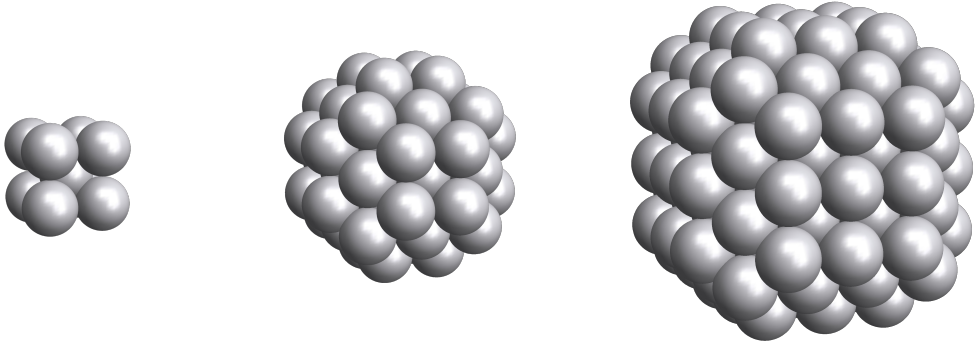

Figure 2: The BCC central 9-atom cluster (left), critical cluster consisting of 5 coordination shells (middle), and the largest cluster consisting of 10 coordination shells (right).

Table 4: Hexagonal close-packed (HCP) shell structure. Table layout is identical to that of Table 2.

| **Coordination shell $n$** | 0 | 1 | 2 | 3 | 4 | 5 | 6 | 7 | 8 | 9 | 10 | 11 |
|---|---|---|---|---|---|---|---|---|---|---|---|---|
| Number of $n^{\text{th}}$ neighbors | 1 | 12 | 6 | 2 | 18 | 12 | 6 | 12 | 12 | 6 | 3 | 12 |
| Total atoms in cluster | 1 | 13 | 19 | 21 | 39 | 51 | 57 | 69 | 81 | 87 | 90 | 102 |
| Cluster radius | 0 | 1 | $\sqrt{2}$ | $2\sqrt{\frac{2}{3}}$ | $\sqrt{3}$ | $\sqrt{\frac{11}{3}}$ | 2 | $\sqrt{5}$ | $\sqrt{\frac{17}{3}}$ | $\sqrt{6}$ | $\sqrt{\frac{19}{3}}$ | $2\sqrt{\frac{5}{3}}$ |

Table 5: Diamond cubic (DC) shell structure. Table layout is identical to that of Table 2.

| **Coordination shell $n$** | 0 | 1 | 2 | 3 | 4 | 5 | 6 | 7 | 8 | 9 | 10 | 11 |
|---|---|---|---|---|---|---|---|---|---|---|---|---|
| Number of $n^{\text{th}}$ neighbors | 0 | 4 | 12 | 12 | 6 | 12 | 24 | 16 | 12 | 24 | 12 | 8 |
| Total atoms in cluster | 1 | 5 | 17 | 29 | 35 | 47 | 71 | 87 | 99 | 123 | 135 | 143 |
| Cluster radius | 0 | 1 | $2\sqrt{\frac{2}{3}}$ | $\sqrt{\frac{11}{3}}$ | $4\sqrt{\frac{1}{3}}$ | $\sqrt{\frac{19}{3}}$ | $2\sqrt{2}$ | 3 | $4\sqrt{\frac{2}{3}}$ | $\sqrt{\frac{35}{3}}$ | $\sqrt{\frac{43}{3}}$ | 4 |

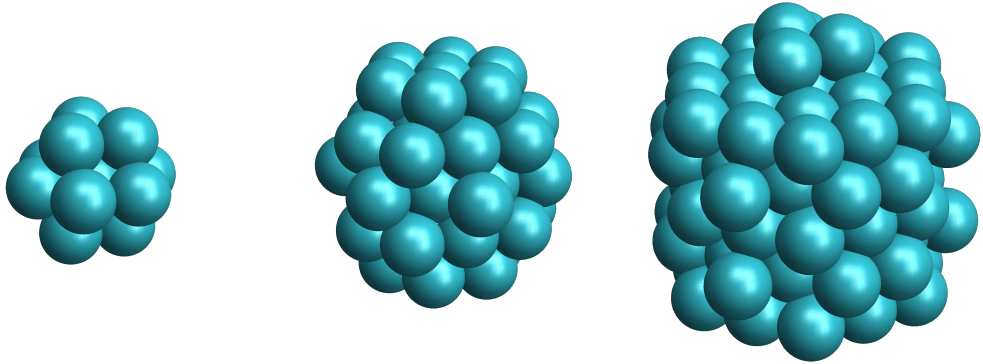

Figure 3: The HCP central 13-atom cluster (left), critical cluster consisting of 5 coordination shells (middle), and the largest cluster consisting of 11 coordination shells (right).

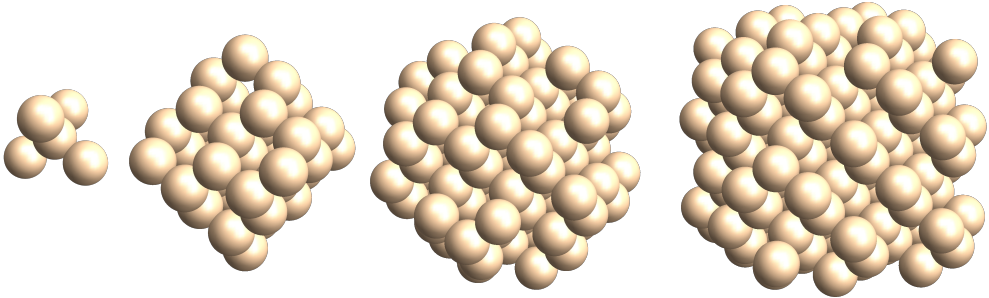

Figure 4: (from the left) The DC central 5-atom cluster, the 4-coordination cluster closing the central atom, the 7-coordination cluster closing the central cluster, and the largest cluster consisting of 11 coordination shells.

# 4  Element data

Data from which the graphs shown in the main text were constructed are provided in Tables 5-8.

Table 6: Perturbation energies $\Delta E$ of the (single) DC cluster investigated, reported in eV.

| Element | Diameter (Å) | $E_f$ | $E_0$ | $\Delta E_1$ | $\Delta E_2$ | $\Delta E_3$ | $\Delta E_4$ | $\Delta E_5$ | $\Delta E_6$ | $\Delta E_7$ | $\Delta E_8$ | $\Delta E_9$ | $\Delta E_{11}$ |
|---|---|---|---|---|---|---|---|---|---|---|---|---|---|
| Si | 2.352 | -5.42 | -7865.72 | -3.05 | -9.01 | -5.66 | -5.55 | -6.09 | -6.71 | -6.67 | -6.12 | -5.91 | -5.79 |

Table 7: Perturbation energies $\Delta E$ of FCC clusters, reported in eV.

| Element | Diameter (Å) | $E_f$ | $E_0$ | $\Delta E_1$ | $\Delta E_2$ | $\Delta E_3$ | $\Delta E_4$ | $\Delta E_5$ | $\Delta E_6$ | $\Delta E_7$ | $\Delta E_9$ |
|---|---|---|---|---|---|---|---|---|---|---|---|
| Al | 2.864 | -3.78 | -6587.00 | -8.80 | -5.86 | -7.20 | - | -4.65 | - | -3.48 | -4.08 |
| Cu | 2.553 | -3.54 | -44611.75 | -3.64 | -5.66 | -3.83 | - | -4.37 | - | -3.56 | -3.56 |
| Rh | 2.758 | -6.78 | -127504.31 | -3.05 | -4.00 | -5.87 | - | -6.32 | - | -7.08 | - |
| Pd | 2.751 | -3.99 | -134362.32 | -1.52 | -2.22 | -4.07 | - | -4.71 | - | -4.88 | - |
| Ag | 2.885 | -3.71 | -141429.58 | -1.38 | -2.13 | -3.83 | - | -3.90 | - | -4.01 | - |

Table 8: Perturbation energies $\Delta E$ of BCC clusters, reported in eV.

| Element | Diameter (Å) | $E_f$ | $E_0$ | $\Delta E_1$ | $\Delta E_2$ | $\Delta E_3$ | $\Delta E_4$ | $\Delta E_5$ | $\Delta E_6$ | $\Delta E_7$ | $\Delta E_8$ | $\Delta E_{10}$ |
|---|---|---|---|---|---|---|---|---|---|---|---|---|
| V | 2.624 | -8.36 | -25665.42 | -3.45 | -4.35 | -7.60 | -8.99 | -8.74 | -8.82 | -9.33 | -9.04 | -8.95 |
| Nb | 2.857 | -10.22 | -102139.80 | -2.77 | -4.06 | -8.71 | -9.39 | -10.93 | -11.02 | -11.00 | -11.06 | - |
| Mo | 2.725 | -10.29 | -108177.90 | -1.57 | -9.08 | -8.29 | -8.82 | -10.16 | -10.15 | -11.51 | -11.06 | - |

Table 9: Perturbation energies $\Delta E$ of HCP clusters, reported in eV.

| Element | Diameter (Å) | $E_f$ | $E_0$ | $\Delta E_1$ | $\Delta E_2$ | $\Delta E_3$ | $\Delta E_4$ | $\Delta E_5$ | $\Delta E_6$ | $\Delta E_7$ | $\Delta E_9$ | $\Delta E_{11}$ |
|---|---|---|---|---|---|---|---|---|---|---|---|---|
| Tc | 2.735 | -11.52 | -114414.78 | -3.95 | -4.87 | - | -9.30 | -10.70 | - | -11.85 | -12.06 | -12.52 |
| Ru | 2.706 | -9.79 | -120855.98 | -4.66 | -5.07 | - | -8.67 | -9.78 | - | -10.84 | -11.10 | -11.37 |

# 5 Dislocation Data

An edge dislocation dipole in Al was generated with the help of molecular dynamics simulations and corresponding clusters were generated from the structure.

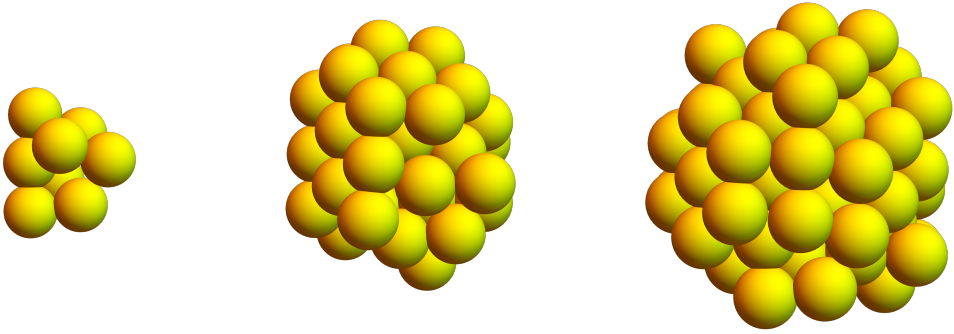

Figure 5: The central 9-atom Al dislcoation cluster (left), critical cluster consisting of 57 atoms (middle), and the largest cluster consisting of 88 atoms (right). These clusters are centered on the non-crystallographic hole corresponding to the dislocation.

Table 10: Perturbation energies $\mathbf{\Delta E}$ to the 9 atom first coordination shell surrounding the dislocation hole at the center in the Al dislcoation clusters Column 1: Number of atoms in the cluster. Coulmn 2: Cluster radii in units of Å. Coulmn 3: per atom energies of central cluster in eV – relative to isolated central cluster.

| Number of atoms in cluster | Cluster radii (Å) | $\mathbf{\Delta E}$ (eV/atom) |
|---|---|---|
| 9 | 3.190 | 0.00 |
| 14 | 3.943 | -0.32 |
| 17 | 4.162 | -0.45 |
| 23 | 4.453 | -0.67 |
| 31 | 4.899 | -0.75 |
| 38 | 5.397 | -0.87 |
| 40 | 5.446 | -0.87 |
| 50 | 5.813 | -0.91 |
| 57 | 6.266 | -0.80 |
| 64 | 6.452 | -0.81 |
| 71 | 6.724 | -0.70 |
| 88 | 7.311 | -0.70 |

# 6 Grain boundary data

Grain boundaries are 2-dimensional defects that are present in real materials. They form the interface that separates the orientation in which atoms are stacked; and alter the microstructure, thus affecting performance. The data represented in graphical form in main text is provided in Table 9 and pictures of important clusters in Figure 5.

Table 11: Perturbation energies $\Delta E$ to the 14 atom first coordination shell surrounding the copper atom at the center in the Σ5 grain boundary clusters Column 1: Number of atoms in the cluster. Coulmn 2: Cluster radii in units of Å. Coulmn 4: per atom energies of central cluster with Cu at center in eV – relative to isolated central cluster. Last row: Extrpolated values for $\Delta E_\infty$.

| Number of atoms in cluster | Cluster radii (Å) | $\Delta E$ with Cu (eV/atom) |
|---|---|---|
| 15 | 2.906 | 0.00 |
| 23 | 4.280 | 0.14 |
| 31 | 4.638 | -0.39 |
| 39 | 4.683 | -0.55 |
| 47 | 4.812 | -0.54 |
| 55 | 5.444 | -0.54 |
| 63 | 5.725 | -0.40 |
| 71 | 5.764 | -0.10 |
| 77 | 5.893 | 0.28 |
| 83 | 6.528 | 0.38 |
| 93 | 6.817 | 0.41 |
| 97 | 6.873 | 0.44 |
| ∞ | ∞ | 0.45 |

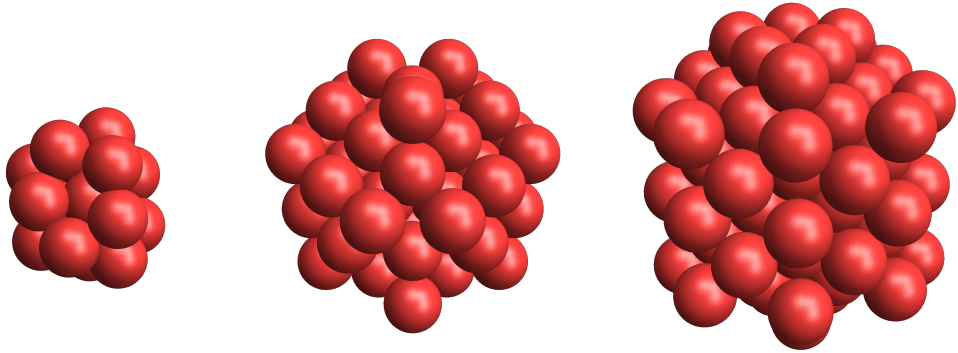

Figure 6: The grain boundary central 15-atom cluster (left), critical cluster consisting of 63 atoms (middle), and the largest cluster consisting of 97 atoms (right).

Table 12: Perturbation energies $\Delta E$ to the 9 atom first coordination shell surrounding the grain boundary hole at the center in the Fe Σ5 grain boundary clusters Column 1: Number of atoms in the cluster. Coulmn 2: Cluster radii in units of Å. Coulmn 4: per atom energies of central cluster in eV – relative to isolated central cluster.

| Number of atoms in cluster | Cluster radii (Å) | $\Delta E$ (eV/atom) |
|---|---|---|
| 9 | 2.606 | -0.26 |
| 11 | 3.436 | -0.26 |
| 15 | 3.507 | -0.24 |
| 23 | 3.968 | -0.11 |
| 29 | 4.026 | -0.17 |
| 33 | 4.240 | -0.16 |
| 37 | 4.386 | -0.16 |
| 41 | 4.551 | -0.03 |
| 47 | 4.965 | 0.01 |
| 53 | 5.431 | 0.00 |



\end{document}